\documentclass[pre,reprint,10pt]{revtex4-1}
\usepackage[left=2.5cm,right=2.5cm,
top=3cm,bottom=3cm]{geometry}
\usepackage[english]{babel}
\usepackage{amsmath,amssymb,amsthm,mathtools}
\usepackage{euscript,mathrsfs}
\usepackage{graphicx}
\usepackage{wrapfig}
\usepackage[dvipsnames]{xcolor}
\usepackage{indentfirst}
\usepackage{amsbsy}
\usepackage{wasysym}
\usepackage{cancel}
\usepackage{verbatim}
\usepackage{empheq}
\usepackage{adjustbox}
\usepackage[pdfencoding=auto]{hyperref}
\definecolor{urlcolor}{HTML}{990000}
\definecolor{linkcolor}{HTML}{005F5F}
\hypersetup{pdfstartview=FitH,  linkcolor=linkcolor,urlcolor=urlcolor, colorlinks=true,citecolor=blue}

\DeclareMathOperator{\im}{\mathrm{Im}\,}

\begin{document}

\title{Information geometry and synchronization phase transition in Kuramoto model}
\author{Artem Alexandrov}
\email{aleksandrov.aa@phystech.edu}
\affiliation{Moscow Institute of Physics and Technology, Dolgoprudny, 141700, Russia}
\affiliation{Laboratory of complex networks, Brain and Consciousness Research Center, Moscow, Russia}

\author{Alexander Gorsky}
\affiliation{Moscow Institute of Physics and Technology, Dolgoprudny, 141700, Russia}
\affiliation{Institute for Information Transmission Problems, Moscow, 127994, Russia}
\affiliation{Laboratory of complex networks, Brain and Consciousness Research Center, Moscow, Russia}

\begin{abstract}
We discuss how the synchronization in Kuramoto model can be treated in terms of information geometry. We argue that the Fisher information is sensitive to synchronization transition, specifically components of Fisher metric diverge at the critical point. Our approach is based on the recently proposed relation between Kuramoto model and geodesics on hyperbolic space.
\end{abstract}

\maketitle

\section{Introduction}
\label{sec:Introduction}

Kuramoto model is the paradigmatic model for synchronization phenomena in non-linear systems. Despite it deceptively simple form, there are still enough white spots, concerning  different aspects of the model. Whereas some of these aspects turn around the synchronization in systems with complicated topology, which were comprehensively described in \cite{rodrigues2016kuramoto}. Another aspects are mostly focused on the rigorous description of the Kuramoto model with all-to-all couplings (complete graph case). It was shown that in case of identical oscillators the Kuramoto model exhibits low-dimensional dynamics. The emergent dimensional reduction occurs due to existence of integrals of motion, initially discovered by trial and error in the papers \cite{Watanabe1993,Watanabe1994}. The investigation of such integrals of motion was done comprehensively in the work \cite{Marvel2009} where it was demonstrated that they emerge due to the invariance of dynamics under the M\"{o}bius transformation. The Kuramoto model dynamics can be represented as the motion on the M\"{o}bius group orbit. Moreover, recently it was noticed that the corresponding low-dimensional dynamics can be described in terms of gradient flows on two dimensional hyperbolic manifold \cite{Chen2017}.

However, the origin of hyperbolic manifold was not established. At first glance, this manifold simply inherits symmetries of M\"{o}bius group and there are no special features. We shall interpret the hyperbolic manifold as induced by probability measures, which is usually refers as statistical manifold (see \cite{Amari2016i} for the review). In particular the AdS$_2$ metric emerges immediately for the shifted Gaussian distribution. In a generic situation the hyperbolic metric emerges as the Fisher metric on the parameter space. The Fisher metric is the probabilistic counterpart of the quantum metric which is familiar in the quantum mechanics where one introduces the so-called complex quantum tensor which has the quantum metric as the real part and the Berry curvature as the imaginary part \cite{Provost_1980,berry1984quantal}. It is built from the quantum wave function depending on some parameter space however its classical analogue exists as well.

In this study we focus on the hyperbolic geometry description of the Kuramoto model and its connection to information geometry. In particular  we explain the hyperbolic geometry involved as the geometry of the statistical manifold. We shall argue that the equation of motion for Kuramoto model has the form of gradient flow for the Kullback-Leibler  divergence. The properties of the induced metric or, in other words, the response of the system at the perturbation of parameters serves as a indicator of the phase transition \cite{Zanardi_2006,Zanardi_2007,you2007fidelity,garnerone2009fidelity}. In particular the diagonal elements of the quantum metric provide the good tool for the search of the critical curves which has been demonstrated in  several examples \cite{Kolodrubetz2013,Kolodrubetz2017,Pandey,buonsante2007ground,dey2012information}. We shall demonstrate in a clear-cut manner that the components of the Fisher information metric identify in a similar manner the critical curve for the phase transition in the Kuramoto-Shakaguchi model. Hence the Fisher information metric yields the proper order parameter for the synchronization phase transition.

The paper is organized as follows. In \autoref{sec:KuramotoBriefReview} we recall the Kuramoto model focusing on its description in terms of the  M\"{o}bius group. The dimensional reduction provides the description of the model in terms of motion of a point at a hyperbolic disc. In \autoref{sec:KuramotoInformGeo} we shall present the key aspects of the information geometry. In \autoref{sec:FisherMetricPhaseTransition} we apply the information geometry to the Kuramoto and Kuramoto-Shakaguchi models. We shall treat the coordinates of point on the disc as the two parameters which the distribution involved depend on. It turns out that the Cauchy distribution governs the Kuramoto model while the von Mises distribution is relevant  for Kuramoto-Shakaguchi model. In both cases we shall demonstrate that the singularity of the Fisher metric coincides with the synchronization transition adding a new example of the relation between the singularities of the Fisher metric and the phase transitions. In \autoref{sec:Conclusion} we formulate the results and the open questions.
The general comments concerning the relations between the distributions and the possible role of the $q$-Gaussian distributions in Kuramoto model can be found in the Appendices.

\section{Brief review of Kuramoto model}
\label{sec:KuramotoBriefReview}

Kuramoto model on a complete graph with identical oscillators has the following equations of motion,
\begin{equation}\label{real-phases-eom}
    \dot{\theta}_i = \omega + \frac{\lambda}{N}\sum_{j=1}^{N}\sin\left(\theta_j-\theta_i\right),
\end{equation}
where $N$ is the number of oscillators, $\theta_i\in [0,2\pi)$, $\lambda>0$ is the coupling constant, and $\omega$ is the oscillator eigenfrequency. The order parameter (synchronization measure) is defined by
\begin{equation}\label{op}
    r(t)=\frac{1}{N}\sum_{j=1}^Nz_j(t),
\end{equation}
where we introduce complex  variable $z_i(t)=\exp(i\theta_i(t))$. The model exhibits so called \emph{synchronization transition} from a given initial state to a synchronized state with the amplitude of order parameter $|r|\rightarrow 1$, which does not depend on time. In terms of particles on the unit circle, it means that in the synchronized state all the particles are concentrated at one point that moves with a certain average frequency.

The equations of motion become
\begin{equation}\label{complex-phase-eom}
    \dot{z}_i = i\omega z_i+\frac{\lambda}{2}\left(r-\overline{r}z_i^2\right).
\end{equation}
The evolution of $z_i=z_i(t)$ can be represented as the action of M\"{o}bius transformation,
\begin{equation}\label{mobius-action}
    z_i(t)=\mathcal{M}(z_i^0),\quad z_i^0\equiv\exp(i\theta_i(t=0)),
\end{equation}
where the parameters of M\"{o}bius transformation depend on time. Following the paper \cite{Chen2017}, this transformation can be parametrized as follows,
\begin{equation}
    \mathcal{M}(z) = \zeta\frac{z-w}{1-\overline{w}z},
\end{equation}
where $\zeta\in\mathbb{C}$, $|\zeta|=1$, $w\in\mathbb{C}$, $|w|<1$. The time derivative of the  phase is
\begin{multline}\label{Mobius}
    \dot{z}_i = -\frac{\dot{w}\zeta}{1-|w|^2}+\left(\dot{\zeta}\overline{\zeta}+\frac{\dot{\overline{w}}w-\dot{w}\overline{w}}{1-|w|^2}\right)z_i + \\ + \frac{\dot{\overline{w}}\overline{\zeta}}{1-|w|^2}z_i^2.
\end{multline}
Matching the RHS of eq.~\eqref{Mobius} and the RHS of eq.~\eqref{complex-phase-eom}, we obtain the system of equations,
\begin{equation}\label{M-low-dim}
\begin{gathered}
    \dot{w} = -\frac{\lambda}{2}\left(1-\left|w\right|^2\right)\overline{\zeta}r,\\ \dot{\zeta}=i\omega\zeta - \frac{\lambda}{2}\left(\overline{w}r-w\overline{r}\zeta^2\right).
\end{gathered}
\end{equation}
The key feature of these equations is that they are decoupled
\begin{multline}\label{eq-decoupling}
    \overline{\zeta}r=\overline{\zeta}\frac{1}{N}\sum_{j=1}^{N}\zeta\frac{z_j^0-w}{1-\overline{w}z_j^0}=\\=\left|\zeta\right|^2\frac{1}{N}\sum_{j=1}^{N}\frac{z_j^0-w}{1-\overline{w}z_j^0}=\frac{1}{N}\sum_{j=1}^{N}\frac{z_j^0-w}{1-\overline{w}z_j^0}.
\end{multline}
hence it means that the equation for $w$ does not feel the variable $\zeta$. This equation is our object of interest and it was shown previously \cite{Chen2017} that it can be rewritten as the gradient flow on the hyperbolic disk $\mathbb{D}=\lbrace z:|z|<1\rbrace$,
\begin{multline}\label{grad-flow-Poisson-kernel}
    \dot{w} =\\= -\frac{\lambda}{2}\left(1-\left|w\right|^2\right)^2\frac{\partial}{\partial\overline{w}}\left\lbrace\frac{1}{N}\sum_{j=1}^{N}\ln P\left(w,z_j^0\right)\right\rbrace,
\end{multline}
where $P=P(w,z_0)$ is the Poisson kernel,
\begin{multline}
    P(w,z_0) = \frac{1-|w|^2}{|w-z_0|^2},\\ z_0\in \mathbb{S}^1,\, w\in \mathbb{D},\,\mathbb{S}^1=\{z:|z|=1\}.
\end{multline}
To determine how the Poisson kernel appears, one should substitute the expression~\eqref{eq-decoupling} into the eq.~\eqref{M-low-dim} and perform integration over $\overline{w}$. In the paper \cite{Chen2017} the authors have already noticed that for $|w|\neq 1$ the fixed point of the eq.~\eqref{grad-flow-Poisson-kernel} coincides with conformal barycenter of $N$ points on $\mathbb{S}^{1}$, which existence and uniqueness are guaranteed \cite{Cantarella2022}. Also one should note that $|w|\rightarrow 1$ limit corresponds to the synchronized state.

Using this representation of the Kuramoto model dynamics, we would like to discuss several features of the model that devoted to the interconnections between information geometry and synchronization.

\section{Information geometry and Kuramoto model}
\label{sec:KuramotoInformGeo}

\subsection{Basic concepts of information geometry}
In this section we discuss the basic concepts of information geometry and develop an information geometry view on the Kuramoto model.  We mainly focus at the continuum  $N\rightarrow\infty$ limit of the Kuramoto model.

To begin with, let us provide some essentials from information geometry. We follow the book \cite{Amari2016i}, where more details can be found. Let us consider the family of probability density functions $p=p(\xi;x)$, where $\xi$ is the vector of parameters and $x$ is the vector of variables.  For all possible values of $\xi$, this family forms the manifold $\mathbb{M}=\{p(\xi;x)\}$, which is called statistical manifold. To determine how two distributions $p(\xi_1;x)$ \& $p(\xi_2;x)$ differ from each other, one can consider the divergence function $D[\xi_1;\xi_2]$. It is possible to define different divergence functions but these functions make sense only if they are invariant and decomposable \cite{amari2009alpha}.

The large class of such divergence functions is called standard $f$-divergencies, that can be represented as
\begin{multline}
    D_f[p;q]=\int dx\,p(x)f\left(\frac{q(x)}{p(x)}\right),\\ f(0)=1,\,f(1)=1,\,f''(1)=1,
\end{multline}
where $f$ is the convex function. For $f(v)=-\ln v$ it is easy to see that corresponding $f$-divergence is nothing more than Kullback-Leibler divergence. Another important type of divergence is so-called $\alpha$-divergence, which is defined by (see \cite{Amari2016i} and \cite{amari2009alpha} for details)
\begin{equation}
    f_{\alpha}(v)=\frac{4}{1-\alpha^2}\left[1-v^{(1+\alpha)/2}\right],\quad \alpha\neq 1.
\end{equation}
The important property of the divergence function is that in case of two close enough points, $\xi_1=\xi+d\xi$ \& $\xi_2=\xi$, it has the following Taylor expansion at point $\xi$,
\begin{equation}
    D[\xi_1;\xi_2]=\frac{1}{2}g_{ij}(\xi)d\xi_id\xi_j+\mathcal{O}(|d\xi|^3),
\end{equation}
where $g_{ij}$ is the metric of manifold $\mathbb{M}$. Any standard $f$-divergence gives the same metric $g_{ij}$, which coincides with the Fisher metric $g^{F}_{ij}$,
\begin{equation}\label{Fisher-metric-def}
    g_{ij}^{\text{F}}=\int dx\,p(x;\xi)\frac{\partial\ln p(x;\xi)}{\partial\xi_i}\frac{\partial\ln p(x;\xi)}{\partial\xi_j}.
\end{equation}
Strictly speaking, to completely describe the statistical manifold $\mathbb{M}$ we need one more object, which is called skewness tensor $T^{\text{F}}$,
\begin{multline}\label{skewness-tensor}
    T_{ijk}^{\text{F}}=\\=\int dx\,p(x;\xi)\frac{\partial\ln p(x;\xi)}{\partial\xi_i}\frac{\partial\ln p(x;\xi)}{\partial\xi_j}\frac{\partial\ln p(x;\xi)}{\partial\xi_k}.
\end{multline}
Finally, the statistical manifold is the triplet $(\mathbb{M},\,g,\,T)$. For a given standard $f$-divergence, the metric tensor $g$ (obtained from the Taylor expansion) coincides with $g^{\text{F}}$, whereas corresponding skewness tensor is given by $T=\alpha T^{\text{F}}$ with $\alpha=2f'''(1)+3$.

The pair of $g^{\text{F}}$ and $T^{\text{F}}$ allows to define so-called $\alpha$-connections on the statistical manifold, that are given by
\begin{equation}
    \Gamma^{\pm\alpha}_{ijk}=\Gamma_{ijk}^0 - \frac{\alpha}{2}T_{ijk}^{\text{F}}.
\end{equation}
The statistical manifold is called $\alpha$-flat if corresponding $\alpha$-Riemannian curvature tensor vanishes everywhere with condition $R_{ijkl}^{\alpha}=R_{ijkl}^{-\alpha}$. Note in contrast with usual Riemannian manifold, the statistical manifold can have non-zero torsion. It was shown that such $\alpha$-flat manifolds has dual affine structure, which gives dual (via Legendre transformation) coordinate system on the statistical manifold. Such manifold is also called dual flat manifold. For dual flat manifold, the metric and skewness tensors are given by
\begin{equation}
    g_{ij}^{\text{F}}=\partial_i\partial_j\psi(\xi),\, T_{ijk}^{\text{F}}=\partial_i\partial_j\partial_k\psi(\xi),\, \partial_i\equiv\frac{\partial}{\partial\xi_i}
\end{equation}
where $\psi(\xi)$ is a convex function. For dual flat manifold, the canonical divergence is so-called the Bregman divergence.

\subsection{Gradient flow on statistical manifold}

For a dual flat manifold, the existence of $\psi(\xi)$ allows to consider the gradient flow with respect to Fisher metric \cite{Nakamura1993,Fujiwara1995},
\begin{equation}\label{stat-manifold-flow}
    \frac{d\xi}{dt} = -\left(g^F\right)^{-1}\partial_{\xi}\psi(\xi).
\end{equation}
Some examples of the gradient flows on statistical manifolds are discussed in  \cite{Nakamura1993}.

The simplest illustrative example of the mentioned gradient flow corresponds to the Gaussian distribution statistical manifold. It is straightforward to check that the statistical manifold formed by Gaussian distributions, $\mathbb{M}_{\text{G}}=\{p(\mu,\sigma;x)|\mu\in\mathbb{R},\,\sigma>0\}$, is $\alpha$-flat for $\alpha=\pm 1$. One can also generalize this statement to the exponential distribution family,
\begin{equation}
    p(\xi;x)=\exp\left[\xi_ig_i(x)+h(x)-\psi(\xi)\right],
\end{equation}
which forms dual flat manifold with canonical coordinate system $\xi$ and the function $\psi=\psi(\xi)$. In case of Gaussian distribution, the coordinate system is $(\xi_1,\xi_2)$ with $\xi_1=\mu/\sigma^2$, $\xi_2=-1/(2\sigma^2)$ and $\psi=\mu^2/(2\sigma^2)+\ln\left(\sqrt{2\pi}\sigma\right)$.
In case of exponential distributions family the Bregman divergence coincides with Kullback-Leibler divergence. The gradient flow for the Gaussian statistical manifold looks as
\begin{equation}\label{Gaussian-grad-flow}
    \dot{\mu}=-\mu;\quad \dot{\sigma}=-\frac{\sigma^2+\mu^2}{2\sigma}.
\end{equation}
Note that this system has the conserved integral of motion $H=\sigma^2/\mu-\mu$, that corresponds to the arc radius.

\subsection{Cauchy distribution}

One can show that family of univariate elliptic distributions forms $\alpha$-flat statistical manifolds \cite{Mitchell1988}. However, there is one orphan distribution in this family: the Cauchy distribution,
\begin{equation}\label{Cauchy-PDF}
    p_{\text{C}}(\gamma,\beta;x)=\frac{1}{\pi}\frac{\gamma}{\gamma^2+(x-\beta)^2},\,\gamma>0,\,\beta\in\mathbb{R},
\end{equation}
which can be written in more convenient for our goals McCullagh's representation \cite{Mccullagh1992},
\begin{equation}\label{McCullagh-Cauchy-PDF}
    p_{\text{C}}(w;x)=\frac{\im w}{\pi|x-w|^2},\quad w=\beta+i\gamma.
\end{equation}
The corresponding Fisher metric looks as
\begin{equation}\label{Cauchy-FisherMetric}
    g_{ij}^{\text{F}}=\frac{1}{2\gamma^2}\begin{pmatrix}1 & 0 \\ 0 & 1\end{pmatrix}
\end{equation}
It is straightforward to verify that for the Cauchy distribution the skewness tensor $T^{\text{F}}$ vanishes everywhere, so we can not find any value of $\alpha$ to obtain dual flat structure at first glance.

However, we can interpret the Cauchy distribution as the member of $q$-Gaussian family, which probability density function is defined as
\begin{equation}\label{qGaussian-definition}
    \mathcal{N}_q(\beta,x)=\frac{\sqrt{\beta}}{C_q}\exp_q\left\{-\beta x^2\right\},\quad \beta>0,
\end{equation}
where $C_q$ is the normalization constant. The Cauchy distribution can be considered as $q$-Gaussian with $q=2$, which was done in work \cite{Nielsen2020} (we discuss some properties of $q$-Gaussians in Appendix),
\begin{equation}\label{qGauss-Cauchy}
    p_{\text{C}}(\gamma,\beta;x)=\left.\mathcal{N}_q\left(\gamma^{-2},(x-\beta)^2\right)\right|_{q=2}.
\end{equation}
Using the definition of $q$-exponent, it is straightforward to verify that the distribution function $p_{\text{C}}(\gamma,\beta;x$) can be also represented as
\begin{equation}\label{Cuachy-qExp}
    p_{\text{C}}(\gamma,\beta;x)=\left.\exp_q\left(\xi_ig_i(x)-\psi(\xi)\right)\right|_{q=2}
\end{equation}
with $\xi_1=2\pi\beta/\gamma$, $\xi_2=-\pi/\gamma$, $g_1(x)=x$, $g_2(x)=x^2$ and the potential function $\psi(\xi)$ is,
\begin{equation}\label{Cauchy-Potential}
    \psi(\xi)=-\frac{\pi^2}{\xi_2}-\frac{\xi_1^2}{4\xi_2}-1.
\end{equation}
Next, it was shown that the following divergence
\begin{multline}\label{BT-Divergence}
    D_{\text{B-T}}[\xi_1;\xi_2]=\\=\left(\int dx\,p(\xi_2;x)^2\right)^{-1}\left[\int dx\,\frac{p(\xi_2;x)^2}{p(\xi_1;x)}-1\right]
\end{multline}
is the Bregman divergence, i.e. describes the dual flat manifold. Such divergence is called Bregman-Tsallis divergence. The metric corresponding to Bregman-Tsallis divergence does not coincide with Fisher metric and can be computed as
\begin{equation}\label{BT-Metric}
    g_{ij}^{q}=\frac{\partial^2\psi(\xi)}{\partial\xi_i\partial\xi_j},
\end{equation}
where $\xi$ corresponds to the canonical coordinate system on dual flat manifold. In case of Cauchy distributions, we have
\begin{equation}\label{BT-Fisher-Relation}
    g_{ij}^{q}=\frac{2\pi}{\gamma}\begin{pmatrix}1 & 0 \\ 0 & 1\end{pmatrix}.
\end{equation}
and it is clear that $g^q$ and $g^{\text{F}}_{\text{C}} $ are related  via conformal transformation. The dual coordinates $\eta_i$ are defined by
\begin{multline}
    \eta_i = \frac{\partial\psi(\xi)}{\partial\xi_i}=\\=\left(-\frac{\xi_1}{2\xi_2},\,\frac{4\pi^2+\xi_1^2}{4\xi_2^2}\right)=(\beta,\,\gamma^2+\beta^2)
\end{multline}
with corresponding dual potential,
\begin{equation}
    \phi(\eta)=1-2\pi\sqrt{\eta_2-\eta_1^2}.
\end{equation}
The gradient flow~\eqref{stat-manifold-flow} for Cauchy distribution looks like
\begin{equation}
    \begin{cases}
        \dot{\xi}_1=\displaystyle\frac{\xi_1}{8\pi^2}\left(4\pi^2+\xi_1^2\right),\\[10pt]
        \dot{\xi}_2=\displaystyle\frac{\xi_2}{8\pi^2}\left(4\pi^2-\xi_1^2\right)
    \end{cases}
    \leftrightarrow
    \begin{cases}
        \dot{\beta}=-\beta,\\[10pt]
        \dot{\gamma}=\displaystyle\frac{\beta^2-\gamma^2}{2\gamma}
    \end{cases}
\end{equation}
Note that these equations are quite similar to the Gaussian case, eq.~\eqref{Gaussian-grad-flow}. Having introduced all relevant concepts from information geometry, we can turn to the interpretation of Kuramoto model in terms of the information geometry.

\section{Fisher metric as the order parameter for synchronization transition}
\label{sec:FisherMetricPhaseTransition}

\subsection{Fisher metric and phase transitions}
Quantum tensor is now the effective tool for analysis of the topological and critical phenomena in the
complicated many-body systems. It is defined as
\begin{equation}
Q_{ij}= \langle\partial_i\Psi|\partial_j \Psi\rangle - \langle\partial_i\Psi| \Psi\rangle\langle\Psi|\partial_j \Psi\rangle
\end{equation}
where $i,j$ indices correspond to the coordinates on the parameter space.
Its real part is the quantum metric while imaginary part is the Berry curvature \cite{Provost_1980,berry1984quantal}
\begin{equation}
    Q_{ij}=g_{ij} + iF_{ij}
\end{equation}
The quantum metric can be thought as a two-point correlator or fidelity whose behavior is expected to quantify the properties of the system \cite{Zanardi_2006,Zanardi_2007,you2007fidelity}. This generic argument can be made more precise if we focus at the geometry of the metric and it was argued that \cite{Zanardi_2006,Zanardi_2007} that the singularity of the metric or Ricci curvature corresponds to the position of the phase transition. The kind of the classification of   induced geometries attributed to the ground state has been found in \cite{Kolodrubetz2013}. The relation between the singularities of the metric and the phase transitions is not a completely rigorous statement but there is convincing list of examples, say \cite{buonsante2007ground,dey2012information} for the Dicke and Hubbard  models. The review of this subject can be found in \cite{Kolodrubetz2017}. The parameters of the Hamiltonian or the momenta can be used to evaluate the components of the quantum metric. More recently it was recognized that the behavior of the metric nearby the singular point can distinguish the integrable or chaotic behavior \cite{Pandey}. The singularities of the different components of the metric provide the information concerning the criticality in the different directions of the parameter space.

The Fisher information metric is parallel to the quantum metric in quantum mechanics when the probability substitutes the squared modulus of the wave function. The probability obeys the Fokker-Planck (FP) equation instead of the Schr\"{o}dinger one hence we shall focus below at solutions to FP equations. The notion of the phase transition in the stochastic problems also requires some care however the relation between the quantum and semiclassical criticalities are seen in the temperature component of the information metric \cite{zanardi2007bures}. It turns out that the synchronization phase transition provides another  clear-cut example when the behavior of the Fisher metric yields the identification of the phase transition in the classical system. In this Section we shall demonstrate that the representation of the Kuramoto model in terms of the distributions does the job. In the pure Kuramoto model we shall see a kind of classical version of quantum phase transition at zero temperature while in
Kuramoto-Sakaguchi model we have an effective temperature due to a noise.

\subsection{Fisher metric in Kuramoto model}

In the paper \cite{Chen2017} the authors have introduced the hyperbolic description of the Kuramoto model. We argue that the hyperbolic space arises as the statistical manifold of wrapped Cauchy distributions. The normalized version of Poisson kernel is given by
\begin{multline}
    p_{\text{wC}}(w;z) = \frac{1}{2\pi}\frac{1-|w|^2}{|w-z|^2},\,|z|=1,\\ \oint_{|z|=1}dz\,\frac{p_{\text{wC}}(w,z)}{iz}=1,
\end{multline}
where wC denotes ``wrapped Cauchy''. It is convenient to use polar representation, $w=re^{i\phi}$ and $z=e^{i\theta}$, which gives
\begin{multline}
    p_{\text{wC}}(r,\phi;\theta)=\frac{1}{2\pi}\frac{1-r^2}{1-2r\cos(\theta-\phi)+r^2},\\ 0\leq r<1,\,\phi=\phi\,\text{mod}\,2\pi
\end{multline}
This is PDF with two parameters $r$ and $\phi$. All such PDFs form manifold $\mathbb{M}_{\text{wC}}$,
\begin{equation*}
    \mathbb{M}_{\text{wC}} = \left\lbrace p_{\text{wC}}(r,\phi;\theta)\,|r\in[0,1),\phi=\phi\,\text{mod}\,2\pi\right\rbrace.
\end{equation*}
and the Fisher metric $g_{ij}^{\text{F}}$ is given by~\eqref{Fisher-metric-def}, which can be represented in more compact form,
\begin{equation*}
    g_{ij}^{\text{F}}=4\int_{-\pi}^{+\pi}d\theta\,\partial_i\sqrt{p_{\text{wC}}(r,\phi;\theta)}\partial_j\sqrt{p_{\text{wC}}(r,\phi;\theta)}.
\end{equation*}
It is straightforward to compute this integral and obtain components of the metric,
\begin{multline}
    g_{rr}^{\text{F}}=\frac{2}{(1-r^2)^2};\quad g_{\phi\phi}^{\text{F}}=\frac{2r^2}{(1-r^2)^2};\\ g_{r\phi}^{\text{F}}=g_{\phi r}^{\text{F}}=0.
\end{multline}
hence, reintroducing $w=re^{i\phi}$, we can write
\begin{equation}
    ds^2_{\text{F}}=\frac{2\left(dr^2+r^2d\phi^2\right)}{(1-r^2)^2}\equiv\frac{2\,dw\,d\overline{w}}{(1-|w|^2)^2},
\end{equation}
which coincides with the hyperbolic disk metrics. Also, the direct computation of~\eqref{skewness-tensor} shows that $T^{\text{F}}\equiv 0$. This is not surprise, because the wrapped Cauchy distribution is simply related to the usual one and the manifold formed by usual Cauchy distributions is nothing more than upper half-plane model of AdS$_2$ space. The wrapped Cauchy distributions also form AdS$_2$ space but in terms of Poincare disk model. Such two models can be mapped to each other via conformal transformation.

We also can represent wrapped Cauchy distribution as $q$-exponent with $q=2$, which can be checked in straightforward way. Computing the canonical coordinates $(\xi_1,\xi_2)$ and then deriving potential function, we can find the metric $g_{ij}^{q}$. In $(w,\,\overline{w})$-coordinates the metric obtained from the corresponding potential function is given by
\begin{equation}
    g_{ij}^{q}=\frac{2\pi}{1-|w|^2}\begin{pmatrix}1 & 0 \\ 0 & 1
    \end{pmatrix}\rightarrow ds_{q}^2=\frac{4\pi\,dw\,d\overline{w}}{1-|w|^2},
\end{equation}
so, the $q$-metric again is conformally equivalent to Fisher metric.

The wrapped Cauchy distribution plays a crucial role in the Kuramoto model dynamics since it is invariant under M\"{o}bius group action. Let $z$ is wrapped random variable with Cauchy distribution, $z\sim p_{\text{wC}}(w;z)$. Let $z'$ is the image of $z$ produced by the M\"{o}bius transformation $\mathcal{M}$, $z'=\mathcal{M}(z)$. It was proven that if $z\sim p_{\text{wC}}(w;z)$, then $z'\sim p_{\text{wC}}(w;z')$. So, wrapped Cauchy distributions are closed with respect action of M\"{o}bius transformation. This fact drives us to conclude that the wrapped Cauchy distribution is a kind of  universal distribution for the Kuramoto model dynamics. Such implication resonates with the role of Ott-Antonsen (O-A) ansatz \cite{Ott2008}.

We would like to emphasize  that Fisher information blows up in the synchronized phase, 
that corresponds to the limit $|w|\rightarrow 1$. In the Kuramoto- Sakaguchi the relation between the singularity of the Fisher metric and the phase transition will be made more transparent.
\subsection{Fisher metric in Kuramoto-Sakaguchi model}
\begin{figure}
    \centering
    \includegraphics[width=\linewidth]{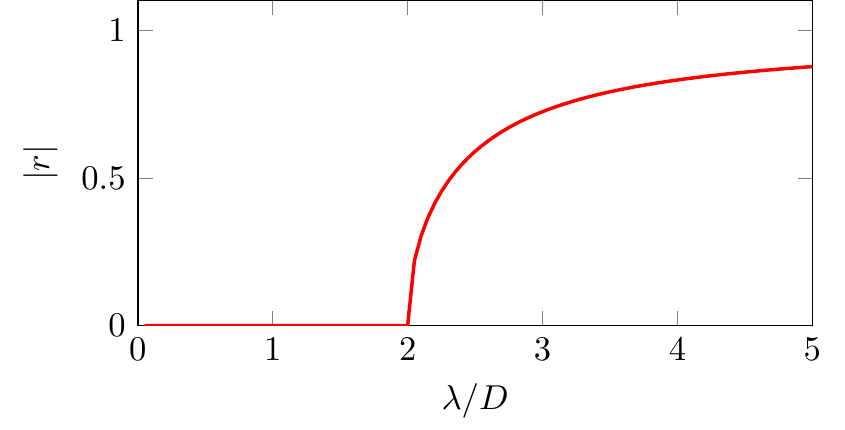}
    \caption{Continuous phase transition in Sakaguchi-Kuramoto model}
    \label{fig:op_noise}
\end{figure}

\begin{figure}
    \centering
    \includegraphics[width=0.75\linewidth]{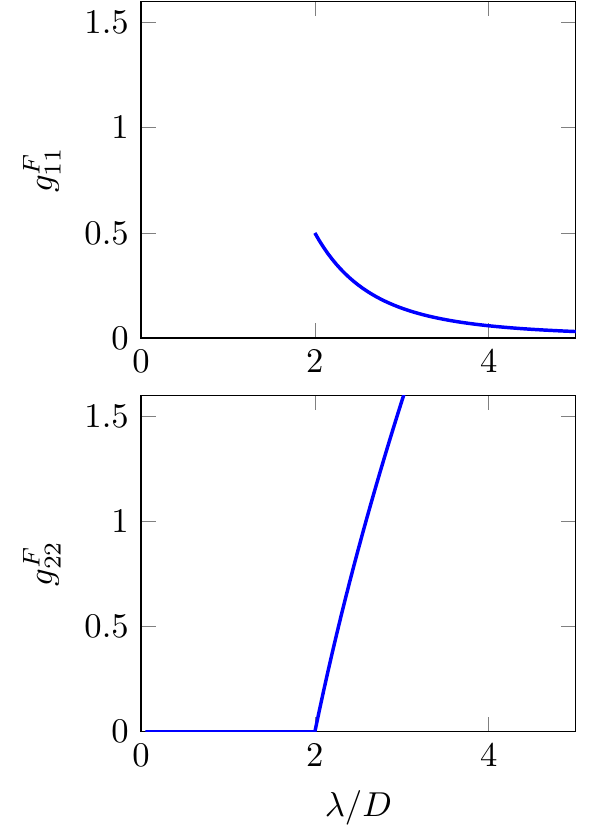}
    \caption{Non-zero components of Fisher metric for Kuramoto-Sakaguchi model as the function of $\lambda/D$ with $\lambda_c=2D$}
    \label{fig:gF_noise}
\end{figure}

\begin{figure}
    \centering
    \includegraphics[width=\linewidth]{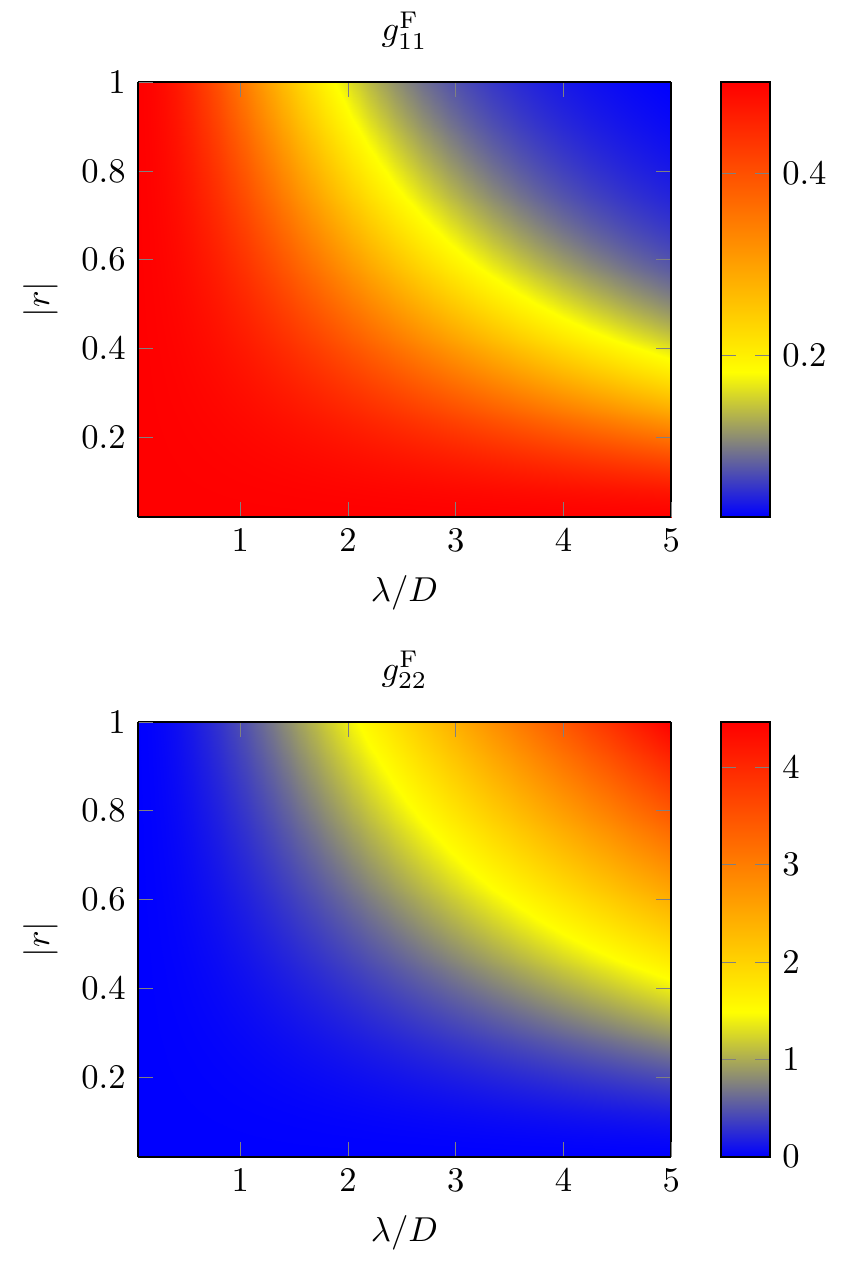}
    \caption{Non-zero components of Fisher metric for Kuramoto-Sakaguchi model as the function of $|r|$ \& $\lambda/D$ with $\lambda_c=2D$}
    \label{fig:gF_surface}
\end{figure}

Consider now the Kuramoto-Sakaguchi model \cite{Sakaguchi1988}, i.e. the Kuramoto model with noise,
\begin{equation}
    \dot{\theta}_i = \omega_i + \frac{\lambda}{N}\sum_{j=1}^{N}\sin\left(\theta_j-\theta_i\right)+\eta_i(t),
\end{equation}
where $\eta_i=\eta_i(t)$ is stochastic term with properties $\langle\eta_i(t)\rangle = 0$, $\langle\eta_i(t)\eta_j(t')\rangle=2D\delta_{ij}\delta(t-t')$, where $D>0$ is noise amplitude. This model was initially considered by Sakaguchi and it was shown that there is the continuous phase transition for the conventional order parameter $r$ with respect to the coupling constant $\lambda$. The critical coupling could be obtained by the analysis of self-consistency equation or by the stability analysis of incoherent state and is given by
\begin{equation}
    \frac{1}{\lambda_c} = \frac{D}{2}\int_{-\infty}^{+\infty}\frac{d\omega\,g(\omega)}{\omega^2+D^2}.
\end{equation}
Both methods deal with the Fokker-Planck equation, which arises in the continuum limit of Kuramoto-Sakaguchi model. In case of identical frequencies, i.e. $g(\omega)=\delta(\omega)$, the critical coupling becomes $\lambda_c=2D$. Moreover, in such case one could easily find the stationary solution of the  Fokker-Planck equation \cite{Goldobin2018}. in rotating reference frame. The stationary solution $\rho_0$ reads as
\begin{multline}
    \rho_0(\theta,\omega) =\\=\frac{1}{2\pi I_0(\lambda|r|/D)}\exp\left\{\frac{\lambda|r|}{D}\cos\left(\arg r - \theta\right)\right\}.
\end{multline}
The self-consistency equation for the Kuramoto-Sakaguchi model is very simple,
\begin{equation}\label{self-consistency-noise}
    |r| = \frac{I_1(\lambda|r|/D)}{I_0(\lambda|r|/D)},
\end{equation}
where it is straightforward to notice that the non-trivial solution exists for $\lambda>\lambda_c=2D$ (see fig.~\ref{fig:op_noise}). This equation is similar with self-consistency equation appeared in XY model and with self-consistency equation appeared in stationary Hamiltonian mean field model. It is not a surprise, because the noise strength $D$ plays role of temperature. In case of non-identical frequencies, the stationary distribution could be found explicitly \cite{Alexandrov2023}, but the Fisher metric can be evaluated only numerically.
The described stationary distribution is nothing more than the von Mises distribution  belonging to the exponential family. In general form, it looks like
\begin{multline}
    p_{\text{vM}}(\kappa,\mu;\theta) = \frac{1}{2\pi I_0(\kappa)}\exp\left\{\kappa\cos(\theta-\mu)\right\},\\ \kappa\geq 0,\,\quad \mu=\mu\,\text{mod}\,2\pi.
\end{multline}
The computation of the Fisher metric for von Mises distribution is straightforward,
\begin{equation}\label{vM-Fisher-metric}
g_{ij}^{\text{F}}=\begin{pmatrix}1-\displaystyle\frac{I_1(\kappa)^2}{I_0(\kappa)^2}-\frac{I_1(\kappa)}{\kappa  I_0(\kappa )} & 0 \\ 0 & \displaystyle\frac{\kappa I_1(\kappa )}{I_0(\kappa )}
\end{pmatrix}.
\end{equation}
where $(i,j)= (\kappa,\mu)$.
Setting $\kappa=\lambda|r|/D$, we obtain the Fisher metric corresponding to the stationary solution of Kuramoto-Sakaguchi model. We can interpret in twofold way: we can set $|r|$ to be a fixed value and then plot the components of metric as the functions of $\lambda$ and $D$ or we can solve the self-consistency equation~\eqref{self-consistency-noise} and find the order parameter $|r|$ as the function of $\lambda$ and $D$ and then consider the components of metric tensor as the functions of $\lambda$ and $D$ only. The second interpretation gives us the following dependencies, fig.~\ref{fig:gF_noise}. The component $g_{11}^{\text{F}}$ does not exist for $\lambda<\lambda_c$, whereas the component $g_{22}^{\text{F}}$ is identically zero for $\lambda<\lambda_c$. The first interpretation says us to plot the components of Fisher metric as functions of order parameter $|r|$ and $\lambda/D$, see fig.~\ref{fig:gF_surface}. The transition point coincides with $\lambda_c=2D$ hence the Fisher metric provides the order parameter for the synchronization transition.

\subsection{Kullback-Leibler divergence in Kuramoto model}

After discussion of the Fisher metric, we would like to notice the following fact: the proposed hyperbolic space description of Kuramoto model dynamics coincides with gradient flow of Kullback-Leibler divergence on the statistical manifold.

Indeed, in the continuum limit, $N\rightarrow\infty$, we deal with initial distribution of phases $f_0=f_0(z)$, $z=e^{i\theta}$. The gradient flow in the limit $N\rightarrow\infty$ becomes
\begin{multline}\label{grad-flow-contin}
    \dot{w} = -\frac{\lambda}{2}\left(1-|w|^2\right)^2\times \\ \times\frac{\partial}{\partial\overline{w}}\oint_{|z|=1}\frac{dz\,f_0(z)\ln p_{\text{wC}}(w;z)}{iz},
\end{multline}
where we rewrite the integral over $\theta$ as the contour integral over the unit circle $|z|=1$ in the complex plane. Our key observation is that the integral in the right hand side of eq.~\eqref{grad-flow-contin} is nothing more than so-called cross entropy of two distributions with PDF $f_0(z)$ and $p_{\text{wC}}(w;z)$,
\begin{equation}
    S_{\text{cross}} = \oint_{|z|=1}\frac{dz\,f_0(z)\ln p_{\text{wC}}(w;z)}{iz}.
\end{equation}
This quantity is tightly connected to the well-known Kullback-Leibler divergence,
\begin{multline}\label{K-L-d}
    S_{\text{K-L}}[f,g] =- \oint_{|z|=1}dz\,\frac{f(z)}{iz}\ln\frac{g(z)}{f(z)}=\\=-S_0[f(z)]-S_{\text{cross}},
\end{multline}
where $f=f(z)$ and $g=g(z)$ are two different PDFs of wrapped distributions and $S_0[f(z)]$ denotes the entropy of distribution $f(z)$. We can notice that gradient with respect to $\overline{w}$ in eq.~\eqref{grad-flow-contin} has the form,
\begin{equation}
    \frac{\partial S_{\text{K-L}}}{\partial\overline{w}}=-\frac{\partial}{\partial\overline{w}}\oint_{|z|=1}\frac{dz\,f(z)\ln g(z)}{iz}.
\end{equation}
So, we conclude that the  gradient flow can be expressed in terms of Kullback-Leibler divergence,
\begin{equation}\label{DKL-equation}
    \dot{w} = +\frac{\lambda}{2}\left(1-\left|w\right|^2\right)^2\frac{\partial S_{\text{K-L}}[f_0,p_{\text{wC}}]}{\partial\overline{w}},
\end{equation}
where $f_0=f_0(z)$ is the initial distribution and $p_{\text{wC}}=p_{\text{wC}}(w;z)$ is the wrapped Cauchy distribution.

As we have mentioned earlier, the Kullback-Leibler divergence in general does not coincide with Bregman divergence. Nevertheless, the eq.~\eqref{DKL-equation} clearly says that the Kuramoto model dynamics can be treated as the gradient flow of the Kullback-Leibler divergence on the statistical manifold formed by wrapped Cauchy distributions. Therefore, to understand such dynamics we can compute the Kullback-Leibler divergence between a given initial distribution $f_0(z)$ and wrapped Cauchy distribution $p_{\text{wC}}(w;z)$. Let us notice that the uniform distribution $p_{\text{U}}(z)=(2\pi)^{-1}$ is nothing more than wrapped Cauchy distribution with $w=0$ and the Dirac-delta distribution is also wrapped Cauchy distribution with $w=1$.

If the initial distribution $f_0(z)$ is uniform, so $f_0(z)=(2\pi)^{-1}$, the Kullback-Leibler divergence it is easy to compute, which gives
\begin{equation}
    S_{\text{K-L}}[f_0,p_{\text{wC}}]=-\ln\left(1-|w|^2\right).
\end{equation}
Substituting this expression into the gradient flow, we find
\begin{equation}
    \dot{w}=\frac{\lambda}{2}\left(1-|w|^2\right)w.
\end{equation}
The obtained equation is interesting for two reasons. First, it is similar to the equation appeared in Ott-Antonsen ansatz. Second, it tells us that in large-$N$ limit the Kuramoto model with all-to-all couplings and identical frequencies is \emph{dual} to the \emph{single} Landau-Stuart oscillator. It is quite straightforward to obtain the solution $w=w(t)$ and than we can see how the Kullback-Leilber divergence evolves in time. Until the transition moment, the divergence is negligibly small (it is quite clear because an instant distribution on the unit circle is still similar to uniform $f_0$). At the transition moment, the Kullback-Leibler starts to grow significantly and finally blows up: the final distribution on the unit circle is drastically different from the uniform, which corresponds to the complete synchronization.

\section{Conclusion}
\label{sec:Conclusion}

In this work we revised the hyperbolic description of the Kuramoto model proposed in \cite{Chen2017} relating it to the statistical manifolds. The key observation of our study  is that the singularity of the Fisher information metric captures the synchronization transition. Namely the  metrics corresponding to the Kuramoto and Kuramoto-Sakaguchi models blow up in the synchronized state. Therefore we extend the list of examples when the phase transitions get identified via the singularities of the Fisher metric. We have focused on the information geometry describing the all-to-all Kuramoto-Shakaguchi model but it would be interesting to recognize the information metric for more general graph architecture like the star when the exact solution is available \cite{Alexandrov2023}. 

The presented approach can be also useful for case with complicated topology, i.e. when an analytical treatment is extremely tough. Based on the fact that Fisher metric has a singularity at the transition point, it seems possible to extract an explicit expressions for model parameters, which raises synchronized state. Moreover, there is no restriction for generalizations of Kuramoto model, which is emphasized by the described example of model with noise. This fact allows to use information geometry framework for generalizations, that includes phase lags, external forces and so on.

We have noticed that the gradient flow in Kuramoto model can be represented as gradient flow on statistical manifold. However, instead of a potential function, one deals with Kullback-Leibler divergence. As was emphasized, Kullback-Leibler divergence is not Bregmann divergence for (wrapped) Cauchy distribution and it is reasonable to represent such flow in terms of Bregmann-Tsallis divergence. But the metric obtained via the potential function of Bregmann-Tsallis divergence does not coincide with Fisher metric (it is conformally equivalent, but different). We address a closer treatment of a connection between synchronization and Bregmann-Tsallis divergence for further research. Next, the discussion of Cauchy distribution as member of $q$-exponential family naturally raises the question concerning wrapped $q$-exponential distributions. As far as we know, the properties of such distributions were not examined. It seems that this research can be interesting in information geometry context.

In our study the coordinates on the  $\mathrm{SL}(2,\mathbb{R})$ orbit play the role of the parameters due to the dimensional reduction of dynamics. Hence to some extend one could be confused why the effective degrees of freedom play the role of the effective parameters. However we know the very precise example of the same  nature --- the duality between the inhomogeneous spin chains and integrable many-body system with long-range interaction of Calogero-Ruijsenaars type which has been formulated in probabilistic terms in \cite{gorsky2022dualities}. It that case the parameters-inhomogeneities in the spin chains get identified with the coordinates of the particles in Calogero model which on the other hand are the coordinates on the group orbit. That is the Fisher metric on the parameters in the spin chain gets mapped into the clear-cut object at the Calogero-Ruijsenaars side. We shall discuss this analogy and relation elsewhere.

The hyperbolic geometry and the large number of degrees of freedom at its boundary invites for the holographic description. However in this case there are  some subtle points since the boundary theory is classical and the equations are of the first order in the time derivatives. Nevertheless the description in terms of the conformal barycenter has a lot in common with the dynamics of  a kind of baryonic vertex since the Poisson kernels are related with the geodesics connecting the points at the boundary and the bulk. We postpone these issues for the separate study. The general discussion concerning the relation of the probabilistic manifold with holography can be found in \cite{Erdmenger_2020}.

The concluding remark concerns the application of the information geometry to the neuroscience. The idea is not new \cite{Amari2016i} however our findings provide the new perspectives for this issue. The Kuramoto model is widely used as the model for the synchronization of the functional connectomes hence one could expect that the properties of the Fisher metric for the Kuramoto model at more general graphs shall provide the new approach for the brain rhythms generation. In particular it would be very interesting to investigate the synchronization of the several functional connectomes via the properties of the Fisher metric and possible dependence of the synchronization of the several functional connectomes on the architecture of the structural connectome.

\section*{Acknowledgements}

We are grateful to A. Polkovnikov for the useful discussions, S. Kato for the discussion on Kato-Jones distribution properties and to F. Nielsen for the comments concerning Cauchy distribution statistical manifold. A.G. thanks IHES and Nordita where the part of the work has been done for the hospitality and support. A.A. thanks grant by Brain Program of the IDEAS Research Center and grant \textnumero 18-12-00434$\Pi$ by Russian Science Foundation. A.G. thanks grant \textnumero 075-15-2020-801 by Ministry of Science and Higher Education of Russian Federation and grant by Basis Foundation.

\appendix

\section{Appendix. On M\"{o}bius transformation and wrapped distributions}
In this Appendix let us describe some facts concerning the wrapped distributions which can be obtained from non-wrapped one via so-called ``compactification''. For random variable $x\in(-\infty,+\infty)$ with probability density function $p(x)$, we introduce the complex variable $z=e^{ix}$ and consider its phase $\theta=\arg z$, $\theta\in(-\pi,+\pi]$, which has \emph{wrapped} probability density function $p_{\text{w}}(\theta)$. The connection between $p(x)$ and its wrapped version is following,
\begin{equation}
    p_{\text{w}}(\theta)=\sum_{k=-\infty}^{+\infty}p(\theta+2\pi k).
\end{equation}
All the common concepts of probability theory work for wrapped distributions as well.

There are four (at least) common wrapped distributions that usually discussed in the context of Kuramoto model: uniform distribution, wrapped Cauchy distribution, von Mises distribution and recently appeared Kato-Jones distribution. The uniform distribution is trivial, the wrapped Cauchy distribution is well-known. The von Mises distribution arises in the Sakaguchi-Kuramoto model as the stationary solution of the corresponding Fokker-Planck equation. The Kato-Jones distribution was introduced in work \cite{Kato2010} and it is M\"{o}bius transformed von Mises distribution. This distribution is invariant under the action of M\"{o}bius group. The mentioned list of distribution is easy to extend. For the sake of completeness, we can also mention cardioid distribution and its M\"{o}bius image, wrapped normal distribution and its M\"{o}bius image.
\begin{figure}[h!]
    \centering
    \includegraphics[width=0.7\linewidth]{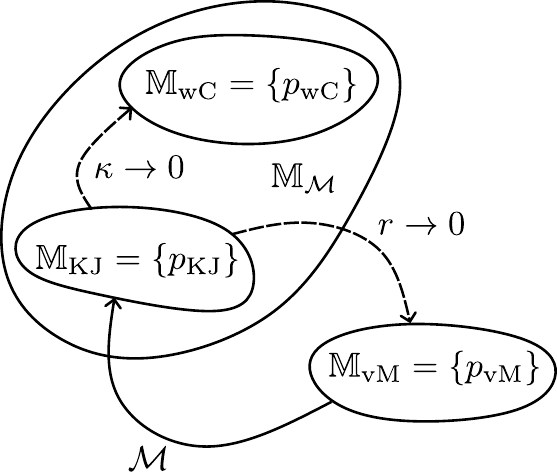}
    \caption{Kuramoto model related statistical manifolds. One can obtain $p_{\text{K-J}}$ via M\"{o}bius transformation $\mathcal{M}$ and then reduce $M_{\text{K-J}}$ to the wrapped Cauchy submanifold $\mathbb{M}_{\text{wC}}$ and the von Mises submanifold $\mathbb{M}_{\text{vM}}$. Both $p_{\text{K-J}}$ and $p_{\text{wC}}$ belong to the M\"{o}bius invariant submanifold $\mathbb{M}_{\text{M}}$, whereas $p_{\text{vM}}$ does not}
    \label{fig:manifolds}
\end{figure}
The invariance under M\"{o}bius group action is crucial in the Kuramoto model. So, here we focus on the triple of wrapped Cauchy distrubition, von Mises distribution and Kato-Jones distribution. We would like to establish the interconnections between members of such triple in the information geometry context. First of all, let us write down the probability density functions for each member of triplet,
\begin{multline}
    p_{\text{wC}}(r,\phi;\theta)=\frac{1}{2\pi}\frac{1-r^2}{1-2r\cos(\theta-\phi)+r^2},\\ 0\leq r<1,\,\phi=\phi\,\text{mod}\,2\pi,
\end{multline}
\begin{multline}
    p_{\text{vM}}(\kappa,\mu;\theta) = \frac{1}{2\pi I_0(\kappa)}\exp\left\{\kappa\cos(\theta-\mu)\right\},\\ \kappa\geq 0,\,\mu=\mu\,\text{mod}\,2\pi,,
\end{multline}
\begin{multline}
    p_{\text{K-J}}(\kappa,r,\nu,\mu;\theta) =\\=\frac{1}{2\pi I_0(\kappa)}\frac{1-r^2}{1-2r\cos(\theta-\mu-\nu)+r^2}\times\\ \times \exp\left\{\frac{\kappa\cos(\mu+\theta)+\kappa r^2\cos(\mu+2\nu-\theta)-2r\kappa\cos\nu}{1-2r\cos(\theta-\mu-\nu)+r^2}\right\}, \\ \kappa\geq 0,\,0\leq r<1,\,\mu=\mu\,\text{mod}\,2\pi,\,\nu=\nu\,\text{mod}\,2\pi
\end{multline}
From these expressions it is straightforward to see the following facts: wrapped Cauchy and von Mises distributions form two dimensional statistical manifolds, the von Mises distribution belongs to exponential family, the Kato-Jones distribution forms four dimensional statistical manifold. The wrapped Cauchy distribution can be derived from the Kato-Jones by setting $\kappa\rightarrow 0$, whereas the von Mises distribution can be derived by setting $r\rightarrow 0$. Therefore, we can treat the statistical manifolds corresponding to the wrapped Cauchy distribution and to the von Mises distribution as the submanifolds of the Kato-Jones statistical manifold. Wrapped distributions $p_{\text{wC}}$ and $p_{\text{K-J}}$ belong to M\"{o}bius invariant statistical manifold $\mathbb{M}_{\mathcal{M}}$, whereas $p_{\text{vM}}$ does not. However, from $p_{\text{K-J}}$ one can obtain $p_{\text{vM}}$ and $p_{\text{wC}}$ taking an appropriate limit, $r\rightarrow 0$ and $\kappa\rightarrow 0$, respectively (see fig.~\ref{fig:manifolds}).

\section{On \texorpdfstring{$q$}{q}-Gaussian distributions in Kuramoto model}

\begin{figure}
    \centering
    \includegraphics[width=\linewidth]{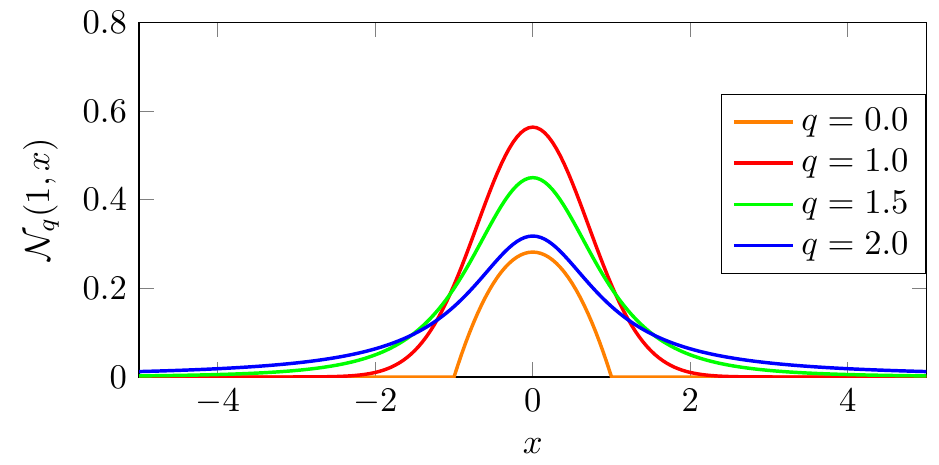}
    \caption{$q$-Gaussians}
    \label{fig:q_Gaussians}
\end{figure}

The Kuramoto model shares some similarities with the famous Hamiltonian mean field (HMF) model \cite{Miritello2009Chaos,Pluchino2006}, which describes $N$ interacting particles on $\mathbb{S}^{1}$ via cosine potential with all-to-all couplings,
\begin{equation}
    H_{\text{HMF}} = \sum_{i=1}^{N}\frac{p_i^2}{2m}+\frac{\lambda}{2N}\sum_{i<j}\cos\left(\theta_j-\theta_i\right),
\end{equation}
where $\lambda$ is the coupling constant. This model is the prototypical for study of systems with long-range interactions (LRI). In the HMF model LRI is caused by all-to-all couplings. The effects of LRI in HMF model are quite well-known: it was shown that LRI causes violent relaxation phenomenon, existence of quasi-stationary states \cite{Yamaguchi2004}. It is worth mentioning that quantum version of HMF model also exhibits such phenomena. One quite interesting consequences of LRI is the appearance of so-called Tsallis $q$-statistics \cite{Tsallis2010}. Due to the presence of LRI, the entropy of a system becomes non-extensive. Tsallis with co-authors developed the appropriate machinery to describe this non-extensivity \cite{Umarov2008}. Roughly speaking, the parameter $q$ measures the non-extensivity in a system. The $q$-statisics deals with so-called $q$-Gaussian distributions (see \cite{Umarov2008} for detailed discussion). Such $q$-Gaussian distributions differ from the usual Gaussian distribution for $q\neq 1$: they have heavier tails (see fig.~\ref{fig:q_Gaussians}) and in the limit $q\rightarrow 1$ one restores usual Gaussian distribution.
Next, in the paper \cite{Tirnakli2007} the authors presented the method to capture correlations between degrees of freedom based on $q$-statistics. Despite the fact that there is no mathematically rigorous derivations of $q$-statistics for a system with LRI, such method allows one to verify (at least numerically) that $q$-statistics takes place. For instance, it was shown that $q$-statistics appears in the HMF model \cite{Pluchino2009,Pluchino2008}.

Based on the idea that HMF model and Kuramoto model are related to each other, in  \cite{Miritello2009CLT} the authors examined fingerprints of $q$-statistics in the Kuramoto model. The key observation was following. From the phases $\theta_i(t)$ one should construct the so-called CLT-variable $y_i$ (see \cite{Tirnakli2007} for details),
\begin{equation}
    y_i=\frac{1}{\sqrt{M}}\sum_{k=1}^{M}\theta_{i}(k\delta),
\end{equation}
where $\delta>0$ is the pre-defined time interval In the Kuramoto model it was  demonstrated that for $\lambda<\lambda_c$ the variables $y_i$ obey $q$-Gaussian distribution with $q\approx 1.7$ , i.e. $q$-CLT appears. The authors considered the case of \emph{non-identical} frequencies and focused on case of uniform distribution and case of Gaussian distribution. In the case of $\lambda>\lambda_c$ the variables $y_i$ were fitted by usual Gaussian distribution.

Having briefly discussed the $q$-Gaussians, we would like to draw attention to the fact that $q$-Gaussian distribution have already appeared in the context of Kuramoto model. Indeed, the continuum limit of Kuramoto model can be described in Ott-Antonsen framework \cite{Ott2008}. In the case of identical oscillators, the distribution function for the continuum limit can be explicitly obtained. This distribution function is nothing more then normalized Poisson kernel, i.e. the wrapped Cauchy distribution. Of course, parameters of the distribution depend on time and initial conditions. As we have shown, the usual Cauchy distribution belongs to $q$-exponential family with $q=2$, so the wrapped Cauchy distribution also inherits properties of $q$-exponential family (wrapped and non-wrapped versions of Cauchy distribution are related to each other via conformal transformation).

\bibliography{references.bib}

\begin{thebibliography}{41}%
\makeatletter
\providecommand \@ifxundefined [1]{%
 \@ifx{#1\undefined}
}%
\providecommand \@ifnum [1]{%
 \ifnum #1\expandafter \@firstoftwo
 \else \expandafter \@secondoftwo
 \fi
}%
\providecommand \@ifx [1]{%
 \ifx #1\expandafter \@firstoftwo
 \else \expandafter \@secondoftwo
 \fi
}%
\providecommand \natexlab [1]{#1}%
\providecommand \enquote  [1]{``#1''}%
\providecommand \bibnamefont  [1]{#1}%
\providecommand \bibfnamefont [1]{#1}%
\providecommand \citenamefont [1]{#1}%
\providecommand \href@noop [0]{\@secondoftwo}%
\providecommand \href [0]{\begingroup \@sanitize@url \@href}%
\providecommand \@href[1]{\@@startlink{#1}\@@href}%
\providecommand \@@href[1]{\endgroup#1\@@endlink}%
\providecommand \@sanitize@url [0]{\catcode `\\12\catcode `\$12\catcode
  `\&12\catcode `\#12\catcode `\^12\catcode `\_12\catcode `\%12\relax}%
\providecommand \@@startlink[1]{}%
\providecommand \@@endlink[0]{}%
\providecommand \url  [0]{\begingroup\@sanitize@url \@url }%
\providecommand \@url [1]{\endgroup\@href {#1}{\urlprefix }}%
\providecommand \urlprefix  [0]{URL }%
\providecommand \Eprint [0]{\href }%
\providecommand \doibase [0]{http://dx.doi.org/}%
\providecommand \selectlanguage [0]{\@gobble}%
\providecommand \bibinfo  [0]{\@secondoftwo}%
\providecommand \bibfield  [0]{\@secondoftwo}%
\providecommand \translation [1]{[#1]}%
\providecommand \BibitemOpen [0]{}%
\providecommand \bibitemStop [0]{}%
\providecommand \bibitemNoStop [0]{.\EOS\space}%
\providecommand \EOS [0]{\spacefactor3000\relax}%
\providecommand \BibitemShut  [1]{\csname bibitem#1\endcsname}%
\let\auto@bib@innerbib\@empty
\bibitem [{\citenamefont {Rodrigues}\ \emph {et~al.}(2016)\citenamefont
  {Rodrigues}, \citenamefont {Peron}, \citenamefont {Ji},\ and\ \citenamefont
  {Kurths}}]{rodrigues2016kuramoto}%
  \BibitemOpen
  \bibfield  {author} {\bibinfo {author} {\bibfnamefont {F.~A.}\ \bibnamefont
  {Rodrigues}}, \bibinfo {author} {\bibfnamefont {T.~K.~D.}\ \bibnamefont
  {Peron}}, \bibinfo {author} {\bibfnamefont {P.}~\bibnamefont {Ji}}, \ and\
  \bibinfo {author} {\bibfnamefont {J.}~\bibnamefont {Kurths}},\ }\href
  {\doibase 10.1016/j.physrep.2015.10.008} {\bibfield  {journal} {\bibinfo
  {journal} {Physics Reports}\ }\textbf {\bibinfo {volume} {610}},\ \bibinfo
  {pages} {1} (\bibinfo {year} {2016})}\BibitemShut {NoStop}%
\bibitem [{\citenamefont {Watanabe}\ and\ \citenamefont
  {Strogatz}(1993)}]{Watanabe1993}%
  \BibitemOpen
  \bibfield  {author} {\bibinfo {author} {\bibfnamefont {S.}~\bibnamefont
  {Watanabe}}\ and\ \bibinfo {author} {\bibfnamefont {S.~H.}\ \bibnamefont
  {Strogatz}},\ }\href {\doibase 10.1103/PhysRevLett.70.2391} {\bibfield
  {journal} {\bibinfo  {journal} {Physical review letters}\ }\textbf {\bibinfo
  {volume} {70}},\ \bibinfo {pages} {2391} (\bibinfo {year}
  {1993})}\BibitemShut {NoStop}%
\bibitem [{\citenamefont {Watanabe}\ and\ \citenamefont
  {Strogatz}(1994)}]{Watanabe1994}%
  \BibitemOpen
  \bibfield  {author} {\bibinfo {author} {\bibfnamefont {S.}~\bibnamefont
  {Watanabe}}\ and\ \bibinfo {author} {\bibfnamefont {S.~H.}\ \bibnamefont
  {Strogatz}},\ }\href {\doibase 10.1016/0167-2789(94)90196-1} {\bibfield
  {journal} {\bibinfo  {journal} {Physica D: Nonlinear Phenomena}\ }\textbf
  {\bibinfo {volume} {74}},\ \bibinfo {pages} {197} (\bibinfo {year}
  {1994})}\BibitemShut {NoStop}%
\bibitem [{\citenamefont {Marvel}\ \emph {et~al.}(2009)\citenamefont {Marvel},
  \citenamefont {Mirollo},\ and\ \citenamefont {Strogatz}}]{Marvel2009}%
  \BibitemOpen
  \bibfield  {author} {\bibinfo {author} {\bibfnamefont {S.~A.}\ \bibnamefont
  {Marvel}}, \bibinfo {author} {\bibfnamefont {R.~E.}\ \bibnamefont {Mirollo}},
  \ and\ \bibinfo {author} {\bibfnamefont {S.~H.}\ \bibnamefont {Strogatz}},\
  }\href {\doibase 10.1063/1.3247089} {\bibfield  {journal} {\bibinfo
  {journal} {Chaos: An Interdisciplinary Journal of Nonlinear Science}\
  }\textbf {\bibinfo {volume} {19}},\ \bibinfo {pages} {043104} (\bibinfo
  {year} {2009})}\BibitemShut {NoStop}%
\bibitem [{\citenamefont {Chen}\ \emph {et~al.}(2017)\citenamefont {Chen},
  \citenamefont {Engelbrecht},\ and\ \citenamefont {Mirollo}}]{Chen2017}%
  \BibitemOpen
  \bibfield  {author} {\bibinfo {author} {\bibfnamefont {B.}~\bibnamefont
  {Chen}}, \bibinfo {author} {\bibfnamefont {J.~R.}\ \bibnamefont
  {Engelbrecht}}, \ and\ \bibinfo {author} {\bibfnamefont {R.}~\bibnamefont
  {Mirollo}},\ }\href {\doibase 10.1088/1751-8121/aa7e39} {\bibfield  {journal}
  {\bibinfo  {journal} {Journal of Physics A: Mathematical and Theoretical}\
  }\textbf {\bibinfo {volume} {50}},\ \bibinfo {pages} {355101} (\bibinfo
  {year} {2017})}\BibitemShut {NoStop}%
\bibitem [{\citenamefont {Amari}(2016)}]{Amari2016i}%
  \BibitemOpen
  \bibfield  {author} {\bibinfo {author} {\bibfnamefont {S.-I.}\ \bibnamefont
  {Amari}},\ }\href {\doibase 10.1007/978-4-431-55978-8} {\emph {\bibinfo
  {title} {Information geometry and its applications}}},\ Vol.\ \bibinfo
  {volume} {194}\ (\bibinfo  {publisher} {Springer},\ \bibinfo {year}
  {2016})\BibitemShut {NoStop}%
\bibitem [{\citenamefont {Provost}\ and\ \citenamefont
  {Vallee}(1980)}]{Provost_1980}%
  \BibitemOpen
  \bibfield  {author} {\bibinfo {author} {\bibfnamefont {J.}~\bibnamefont
  {Provost}}\ and\ \bibinfo {author} {\bibfnamefont {G.}~\bibnamefont
  {Vallee}},\ }\href {\doibase 10.1007/BF02193559} {\bibfield  {journal}
  {\bibinfo  {journal} {Communications in Mathematical Physics}\ }\textbf
  {\bibinfo {volume} {76}},\ \bibinfo {pages} {289} (\bibinfo {year}
  {1980})}\BibitemShut {NoStop}%
\bibitem [{\citenamefont {Berry}(1984)}]{berry1984quantal}%
  \BibitemOpen
  \bibfield  {author} {\bibinfo {author} {\bibfnamefont {M.~V.}\ \bibnamefont
  {Berry}},\ }\href {\doibase 10.1098/rspa.1984.0023} {\bibfield  {journal}
  {\bibinfo  {journal} {Proceedings of the Royal Society of London. A.
  Mathematical and Physical Sciences}\ }\textbf {\bibinfo {volume} {392}},\
  \bibinfo {pages} {45} (\bibinfo {year} {1984})}\BibitemShut {NoStop}%
\bibitem [{\citenamefont {Zanardi}\ and\ \citenamefont
  {Paunkovi\ifmmode~\acute{c}\else \'{c}\fi{}}(2006)}]{Zanardi_2006}%
  \BibitemOpen
  \bibfield  {author} {\bibinfo {author} {\bibfnamefont {P.}~\bibnamefont
  {Zanardi}}\ and\ \bibinfo {author} {\bibfnamefont {N.}~\bibnamefont
  {Paunkovi\ifmmode~\acute{c}\else \'{c}\fi{}}},\ }\href {\doibase
  10.1103/PhysRevE.74.031123} {\bibfield  {journal} {\bibinfo  {journal} {Phys.
  Rev. E}\ }\textbf {\bibinfo {volume} {74}},\ \bibinfo {pages} {031123}
  (\bibinfo {year} {2006})}\BibitemShut {NoStop}%
\bibitem [{\citenamefont {Zanardi}\ \emph
  {et~al.}(2007{\natexlab{a}})\citenamefont {Zanardi}, \citenamefont {Giorda},\
  and\ \citenamefont {Cozzini}}]{Zanardi_2007}%
  \BibitemOpen
  \bibfield  {author} {\bibinfo {author} {\bibfnamefont {P.}~\bibnamefont
  {Zanardi}}, \bibinfo {author} {\bibfnamefont {P.}~\bibnamefont {Giorda}}, \
  and\ \bibinfo {author} {\bibfnamefont {M.}~\bibnamefont {Cozzini}},\ }\href
  {\doibase 10.1103/PhysRevLett.99.100603} {\bibfield  {journal} {\bibinfo
  {journal} {Phys. Rev. Lett.}\ }\textbf {\bibinfo {volume} {99}},\ \bibinfo
  {pages} {100603} (\bibinfo {year} {2007}{\natexlab{a}})}\BibitemShut
  {NoStop}%
\bibitem [{\citenamefont {You}\ \emph {et~al.}(2007)\citenamefont {You},
  \citenamefont {Li},\ and\ \citenamefont {Gu}}]{you2007fidelity}%
  \BibitemOpen
  \bibfield  {author} {\bibinfo {author} {\bibfnamefont {W.-L.}\ \bibnamefont
  {You}}, \bibinfo {author} {\bibfnamefont {Y.-W.}\ \bibnamefont {Li}}, \ and\
  \bibinfo {author} {\bibfnamefont {S.-J.}\ \bibnamefont {Gu}},\ }\href
  {\doibase 10.1103/PhysRevE.76.022101} {\bibfield  {journal} {\bibinfo
  {journal} {Physical Review E}\ }\textbf {\bibinfo {volume} {76}},\ \bibinfo
  {pages} {022101} (\bibinfo {year} {2007})}\BibitemShut {NoStop}%
\bibitem [{\citenamefont {Garnerone}\ \emph {et~al.}(2009)\citenamefont
  {Garnerone}, \citenamefont {Abasto}, \citenamefont {Haas},\ and\
  \citenamefont {Zanardi}}]{garnerone2009fidelity}%
  \BibitemOpen
  \bibfield  {author} {\bibinfo {author} {\bibfnamefont {S.}~\bibnamefont
  {Garnerone}}, \bibinfo {author} {\bibfnamefont {D.}~\bibnamefont {Abasto}},
  \bibinfo {author} {\bibfnamefont {S.}~\bibnamefont {Haas}}, \ and\ \bibinfo
  {author} {\bibfnamefont {P.}~\bibnamefont {Zanardi}},\ }\href {\doibase
  10.1103/PhysRevA.79.032302} {\bibfield  {journal} {\bibinfo  {journal}
  {Physical Review A}\ }\textbf {\bibinfo {volume} {79}},\ \bibinfo {pages}
  {032302} (\bibinfo {year} {2009})}\BibitemShut {NoStop}%
\bibitem [{\citenamefont {Kolodrubetz}\ \emph {et~al.}(2013)\citenamefont
  {Kolodrubetz}, \citenamefont {Gritsev},\ and\ \citenamefont
  {Polkovnikov}}]{Kolodrubetz2013}%
  \BibitemOpen
  \bibfield  {author} {\bibinfo {author} {\bibfnamefont {M.}~\bibnamefont
  {Kolodrubetz}}, \bibinfo {author} {\bibfnamefont {V.}~\bibnamefont
  {Gritsev}}, \ and\ \bibinfo {author} {\bibfnamefont {A.}~\bibnamefont
  {Polkovnikov}},\ }\href {\doibase 10.1103/PhysRevB.88.064304} {\bibfield
  {journal} {\bibinfo  {journal} {Physical Review B}\ }\textbf {\bibinfo
  {volume} {88}},\ \bibinfo {pages} {064304} (\bibinfo {year}
  {2013})}\BibitemShut {NoStop}%
\bibitem [{\citenamefont {Kolodrubetz}\ \emph {et~al.}(2017)\citenamefont
  {Kolodrubetz}, \citenamefont {Sels}, \citenamefont {Mehta},\ and\
  \citenamefont {Polkovnikov}}]{Kolodrubetz2017}%
  \BibitemOpen
  \bibfield  {author} {\bibinfo {author} {\bibfnamefont {M.}~\bibnamefont
  {Kolodrubetz}}, \bibinfo {author} {\bibfnamefont {D.}~\bibnamefont {Sels}},
  \bibinfo {author} {\bibfnamefont {P.}~\bibnamefont {Mehta}}, \ and\ \bibinfo
  {author} {\bibfnamefont {A.}~\bibnamefont {Polkovnikov}},\ }\href {\doibase
  10.1016/j.physrep.2017.07.001} {\bibfield  {journal} {\bibinfo  {journal}
  {Physics Reports}\ }\textbf {\bibinfo {volume} {697}},\ \bibinfo {pages} {1}
  (\bibinfo {year} {2017})}\BibitemShut {NoStop}%
\bibitem [{\citenamefont {Pandey}\ \emph {et~al.}(2020)\citenamefont {Pandey},
  \citenamefont {Claeys}, \citenamefont {Campbell}, \citenamefont
  {Polkovnikov},\ and\ \citenamefont {Sels}}]{Pandey}%
  \BibitemOpen
  \bibfield  {author} {\bibinfo {author} {\bibfnamefont {M.}~\bibnamefont
  {Pandey}}, \bibinfo {author} {\bibfnamefont {P.~W.}\ \bibnamefont {Claeys}},
  \bibinfo {author} {\bibfnamefont {D.~K.}\ \bibnamefont {Campbell}}, \bibinfo
  {author} {\bibfnamefont {A.}~\bibnamefont {Polkovnikov}}, \ and\ \bibinfo
  {author} {\bibfnamefont {D.}~\bibnamefont {Sels}},\ }\href {\doibase
  10.1103/PhysRevX.10.041017} {\bibfield  {journal} {\bibinfo  {journal} {Phys.
  Rev. X}\ }\textbf {\bibinfo {volume} {10}},\ \bibinfo {pages} {041017}
  (\bibinfo {year} {2020})}\BibitemShut {NoStop}%
\bibitem [{\citenamefont {Buonsante}\ and\ \citenamefont
  {Vezzani}(2007)}]{buonsante2007ground}%
  \BibitemOpen
  \bibfield  {author} {\bibinfo {author} {\bibfnamefont {P.}~\bibnamefont
  {Buonsante}}\ and\ \bibinfo {author} {\bibfnamefont {A.}~\bibnamefont
  {Vezzani}},\ }\href {\doibase 10.1103/PhysRevLett.98.110601} {\bibfield
  {journal} {\bibinfo  {journal} {Physical review letters}\ }\textbf {\bibinfo
  {volume} {98}},\ \bibinfo {pages} {110601} (\bibinfo {year}
  {2007})}\BibitemShut {NoStop}%
\bibitem [{\citenamefont {Dey}\ \emph {et~al.}(2012)\citenamefont {Dey},
  \citenamefont {Mahapatra}, \citenamefont {Roy},\ and\ \citenamefont
  {Sarkar}}]{dey2012information}%
  \BibitemOpen
  \bibfield  {author} {\bibinfo {author} {\bibfnamefont {A.}~\bibnamefont
  {Dey}}, \bibinfo {author} {\bibfnamefont {S.}~\bibnamefont {Mahapatra}},
  \bibinfo {author} {\bibfnamefont {P.}~\bibnamefont {Roy}}, \ and\ \bibinfo
  {author} {\bibfnamefont {T.}~\bibnamefont {Sarkar}},\ }\href {\doibase
  10.1103/PhysRevE.86.031137} {\bibfield  {journal} {\bibinfo  {journal}
  {Physical Review E}\ }\textbf {\bibinfo {volume} {86}},\ \bibinfo {pages}
  {031137} (\bibinfo {year} {2012})}\BibitemShut {NoStop}%
\bibitem [{\citenamefont {Cantarella}\ and\ \citenamefont
  {Schumacher}(2022)}]{Cantarella2022}%
  \BibitemOpen
  \bibfield  {author} {\bibinfo {author} {\bibfnamefont {J.}~\bibnamefont
  {Cantarella}}\ and\ \bibinfo {author} {\bibfnamefont {H.}~\bibnamefont
  {Schumacher}},\ }\href {\doibase 10.1137/21M1449282} {\bibfield  {journal}
  {\bibinfo  {journal} {SIAM Journal on Applied Algebra and Geometry}\ }\textbf
  {\bibinfo {volume} {6}},\ \bibinfo {pages} {503} (\bibinfo {year}
  {2022})}\BibitemShut {NoStop}%
\bibitem [{\citenamefont {Amari}(2009)}]{amari2009alpha}%
  \BibitemOpen
  \bibfield  {author} {\bibinfo {author} {\bibfnamefont {S.-I.}\ \bibnamefont
  {Amari}},\ }\href {\doibase 10.1109/TIT.2009.2030485} {\bibfield  {journal}
  {\bibinfo  {journal} {IEEE Transactions on Information Theory}\ }\textbf
  {\bibinfo {volume} {55}},\ \bibinfo {pages} {4925} (\bibinfo {year}
  {2009})}\BibitemShut {NoStop}%
\bibitem [{\citenamefont {Nakamura}(1993)}]{Nakamura1993}%
  \BibitemOpen
  \bibfield  {author} {\bibinfo {author} {\bibfnamefont {Y.}~\bibnamefont
  {Nakamura}},\ }\href {\doibase 10.1007/BF03167571} {\bibfield  {journal}
  {\bibinfo  {journal} {Japan journal of industrial and applied mathematics}\
  }\textbf {\bibinfo {volume} {10}},\ \bibinfo {pages} {179} (\bibinfo {year}
  {1993})}\BibitemShut {NoStop}%
\bibitem [{\citenamefont {Fujiwara}\ and\ \citenamefont
  {Amari}(1995)}]{Fujiwara1995}%
  \BibitemOpen
  \bibfield  {author} {\bibinfo {author} {\bibfnamefont {A.}~\bibnamefont
  {Fujiwara}}\ and\ \bibinfo {author} {\bibfnamefont {S.-i.}\ \bibnamefont
  {Amari}},\ }\href {\doibase 10.1016/0167-2789(94)00175-P} {\bibfield
  {journal} {\bibinfo  {journal} {Physica D: Nonlinear Phenomena}\ }\textbf
  {\bibinfo {volume} {80}},\ \bibinfo {pages} {317} (\bibinfo {year}
  {1995})}\BibitemShut {NoStop}%
\bibitem [{\citenamefont {Mitchell}(1988)}]{Mitchell1988}%
  \BibitemOpen
  \bibfield  {author} {\bibinfo {author} {\bibfnamefont {A.~F.}\ \bibnamefont
  {Mitchell}},\ }\href {\doibase 10.2307/1403358} {\bibfield  {journal}
  {\bibinfo  {journal} {International Statistical Review/Revue Internationale
  de Statistique}\ ,\ \bibinfo {pages} {1}} (\bibinfo {year}
  {1988})}\BibitemShut {NoStop}%
\bibitem [{\citenamefont {McCullagh}(1992)}]{Mccullagh1992}%
  \BibitemOpen
  \bibfield  {author} {\bibinfo {author} {\bibfnamefont {P.}~\bibnamefont
  {McCullagh}},\ }\href {\doibase 10.1093/biomet/79.2.247} {\bibfield
  {journal} {\bibinfo  {journal} {Biometrika}\ }\textbf {\bibinfo {volume}
  {79}},\ \bibinfo {pages} {247} (\bibinfo {year} {1992})}\BibitemShut
  {NoStop}%
\bibitem [{\citenamefont {Nielsen}(2020)}]{Nielsen2020}%
  \BibitemOpen
  \bibfield  {author} {\bibinfo {author} {\bibfnamefont {F.}~\bibnamefont
  {Nielsen}},\ }\href {\doibase 10.3390/e22070713} {\bibfield  {journal}
  {\bibinfo  {journal} {Entropy}\ }\textbf {\bibinfo {volume} {22}},\ \bibinfo
  {pages} {713} (\bibinfo {year} {2020})}\BibitemShut {NoStop}%
\bibitem [{\citenamefont {Zanardi}\ \emph
  {et~al.}(2007{\natexlab{b}})\citenamefont {Zanardi}, \citenamefont {Venuti},\
  and\ \citenamefont {Giorda}}]{zanardi2007bures}%
  \BibitemOpen
  \bibfield  {author} {\bibinfo {author} {\bibfnamefont {P.}~\bibnamefont
  {Zanardi}}, \bibinfo {author} {\bibfnamefont {L.~C.}\ \bibnamefont {Venuti}},
  \ and\ \bibinfo {author} {\bibfnamefont {P.}~\bibnamefont {Giorda}},\ }\href
  {\doibase 10.1103/PhysRevA.76.062318} {\bibfield  {journal} {\bibinfo
  {journal} {Physical Review A}\ }\textbf {\bibinfo {volume} {76}},\ \bibinfo
  {pages} {062318} (\bibinfo {year} {2007}{\natexlab{b}})}\BibitemShut
  {NoStop}%
\bibitem [{\citenamefont {Ott}\ and\ \citenamefont {Antonsen}(2008)}]{Ott2008}%
  \BibitemOpen
  \bibfield  {author} {\bibinfo {author} {\bibfnamefont {E.}~\bibnamefont
  {Ott}}\ and\ \bibinfo {author} {\bibfnamefont {T.~M.}\ \bibnamefont
  {Antonsen}},\ }\href {\doibase 10.1063/1.2930766} {\bibfield  {journal}
  {\bibinfo  {journal} {Chaos: An Interdisciplinary Journal of Nonlinear
  Science}\ }\textbf {\bibinfo {volume} {18}},\ \bibinfo {pages} {037113}
  (\bibinfo {year} {2008})}\BibitemShut {NoStop}%
\bibitem [{\citenamefont {Sakaguchi}(1988)}]{Sakaguchi1988}%
  \BibitemOpen
  \bibfield  {author} {\bibinfo {author} {\bibfnamefont {H.}~\bibnamefont
  {Sakaguchi}},\ }\href {\doibase 10.1143/PTP.79.39} {\bibfield  {journal}
  {\bibinfo  {journal} {Progress of theoretical physics}\ }\textbf {\bibinfo
  {volume} {79}},\ \bibinfo {pages} {39} (\bibinfo {year} {1988})}\BibitemShut
  {NoStop}%
\bibitem [{\citenamefont {Goldobin}\ \emph {et~al.}(2018)\citenamefont
  {Goldobin}, \citenamefont {Tyulkina}, \citenamefont {Klimenko},\ and\
  \citenamefont {Pikovsky}}]{Goldobin2018}%
  \BibitemOpen
  \bibfield  {author} {\bibinfo {author} {\bibfnamefont {D.~S.}\ \bibnamefont
  {Goldobin}}, \bibinfo {author} {\bibfnamefont {I.~V.}\ \bibnamefont
  {Tyulkina}}, \bibinfo {author} {\bibfnamefont {L.~S.}\ \bibnamefont
  {Klimenko}}, \ and\ \bibinfo {author} {\bibfnamefont {A.}~\bibnamefont
  {Pikovsky}},\ }\href {\doibase 10.1063/1.5053576} {\bibfield  {journal}
  {\bibinfo  {journal} {Chaos: An Interdisciplinary Journal of Nonlinear
  Science}\ }\textbf {\bibinfo {volume} {28}},\ \bibinfo {pages} {101101}
  (\bibinfo {year} {2018})}\BibitemShut {NoStop}%
\bibitem [{\citenamefont {Alexandrov}(2023)}]{Alexandrov2023}%
  \BibitemOpen
  \bibfield  {author} {\bibinfo {author} {\bibfnamefont {A.}~\bibnamefont
  {Alexandrov}},\ }\href {\doibase 10.1016/j.chaos.2022.113056} {\bibfield
  {journal} {\bibinfo  {journal} {Chaos, Solitons \& Fractals}\ }\textbf
  {\bibinfo {volume} {167}},\ \bibinfo {pages} {113056} (\bibinfo {year}
  {2023})}\BibitemShut {NoStop}%
\bibitem [{\citenamefont {Gorsky}\ \emph {et~al.}(2022)\citenamefont {Gorsky},
  \citenamefont {Vasilyev},\ and\ \citenamefont {Zotov}}]{gorsky2022dualities}%
  \BibitemOpen
  \bibfield  {author} {\bibinfo {author} {\bibfnamefont {A.}~\bibnamefont
  {Gorsky}}, \bibinfo {author} {\bibfnamefont {M.}~\bibnamefont {Vasilyev}}, \
  and\ \bibinfo {author} {\bibfnamefont {A.}~\bibnamefont {Zotov}},\ }\href
  {\doibase 10.1007/JHEP04(2022)159} {\bibfield  {journal} {\bibinfo  {journal}
  {Journal of High Energy Physics}\ }\textbf {\bibinfo {volume} {2022}},\
  \bibinfo {pages} {1} (\bibinfo {year} {2022})}\BibitemShut {NoStop}%
\bibitem [{\citenamefont {Erdmenger}\ \emph {et~al.}(2020)\citenamefont
  {Erdmenger}, \citenamefont {Grosvenor},\ and\ \citenamefont
  {Jefferson}}]{Erdmenger_2020}%
  \BibitemOpen
  \bibfield  {author} {\bibinfo {author} {\bibfnamefont {J.}~\bibnamefont
  {Erdmenger}}, \bibinfo {author} {\bibfnamefont {K.}~\bibnamefont
  {Grosvenor}}, \ and\ \bibinfo {author} {\bibfnamefont {R.}~\bibnamefont
  {Jefferson}},\ }\href {\doibase 10.21468/SciPostPhys.8.5.073} {\bibfield
  {journal} {\bibinfo  {journal} {SciPost Physics}\ }\textbf {\bibinfo {volume}
  {8}},\ \bibinfo {pages} {073} (\bibinfo {year} {2020})}\BibitemShut {NoStop}%
\bibitem [{\citenamefont {Kato}\ and\ \citenamefont {Jones}(2010)}]{Kato2010}%
  \BibitemOpen
  \bibfield  {author} {\bibinfo {author} {\bibfnamefont {S.}~\bibnamefont
  {Kato}}\ and\ \bibinfo {author} {\bibfnamefont {M.}~\bibnamefont {Jones}},\
  }\href {\doibase 10.1198/jasa.2009.tm08313} {\bibfield  {journal} {\bibinfo
  {journal} {Journal of the American Statistical Association}\ }\textbf
  {\bibinfo {volume} {105}},\ \bibinfo {pages} {249} (\bibinfo {year}
  {2010})}\BibitemShut {NoStop}%
\bibitem [{\citenamefont {Miritello}\ \emph
  {et~al.}(2009{\natexlab{a}})\citenamefont {Miritello}, \citenamefont
  {Pluchino},\ and\ \citenamefont {Rapisarda}}]{Miritello2009Chaos}%
  \BibitemOpen
  \bibfield  {author} {\bibinfo {author} {\bibfnamefont {G.}~\bibnamefont
  {Miritello}}, \bibinfo {author} {\bibfnamefont {A.}~\bibnamefont {Pluchino}},
  \ and\ \bibinfo {author} {\bibfnamefont {A.}~\bibnamefont {Rapisarda}},\
  }\href {\doibase 10.1209/0295-5075/85/10007} {\bibfield  {journal} {\bibinfo
  {journal} {EPL (Europhysics Letters)}\ }\textbf {\bibinfo {volume} {85}},\
  \bibinfo {pages} {10007} (\bibinfo {year} {2009}{\natexlab{a}})}\BibitemShut
  {NoStop}%
\bibitem [{\citenamefont {Pluchino}\ and\ \citenamefont
  {Rapisarda}(2006)}]{Pluchino2006}%
  \BibitemOpen
  \bibfield  {author} {\bibinfo {author} {\bibfnamefont {A.}~\bibnamefont
  {Pluchino}}\ and\ \bibinfo {author} {\bibfnamefont {A.}~\bibnamefont
  {Rapisarda}},\ }\href {\doibase 10.1016/j.physa.2006.01.039} {\bibfield
  {journal} {\bibinfo  {journal} {Physica A: Statistical Mechanics and its
  Applications}\ }\textbf {\bibinfo {volume} {365}},\ \bibinfo {pages} {184}
  (\bibinfo {year} {2006})}\BibitemShut {NoStop}%
\bibitem [{\citenamefont {Yamaguchi}\ \emph {et~al.}(2004)\citenamefont
  {Yamaguchi}, \citenamefont {Barr{\'e}}, \citenamefont {Bouchet},
  \citenamefont {Dauxois},\ and\ \citenamefont {Ruffo}}]{Yamaguchi2004}%
  \BibitemOpen
  \bibfield  {author} {\bibinfo {author} {\bibfnamefont {Y.~Y.}\ \bibnamefont
  {Yamaguchi}}, \bibinfo {author} {\bibfnamefont {J.}~\bibnamefont
  {Barr{\'e}}}, \bibinfo {author} {\bibfnamefont {F.}~\bibnamefont {Bouchet}},
  \bibinfo {author} {\bibfnamefont {T.}~\bibnamefont {Dauxois}}, \ and\
  \bibinfo {author} {\bibfnamefont {S.}~\bibnamefont {Ruffo}},\ }\href
  {\doibase 10.1016/j.physa.2004.01.041} {\bibfield  {journal} {\bibinfo
  {journal} {Physica A: Statistical Mechanics and its Applications}\ }\textbf
  {\bibinfo {volume} {337}},\ \bibinfo {pages} {36} (\bibinfo {year}
  {2004})}\BibitemShut {NoStop}%
\bibitem [{\citenamefont {Tsallis}\ and\ \citenamefont
  {Tirnakli}(2010)}]{Tsallis2010}%
  \BibitemOpen
  \bibfield  {author} {\bibinfo {author} {\bibfnamefont {C.}~\bibnamefont
  {Tsallis}}\ and\ \bibinfo {author} {\bibfnamefont {U.}~\bibnamefont
  {Tirnakli}},\ }in\ \href@noop {} {\emph {\bibinfo {booktitle} {Journal of
  Physics: Conference Series}}}\ (\bibinfo {organization} {IOP Publishing},\
  \bibinfo {year} {2010})\BibitemShut {NoStop}%
\bibitem [{\citenamefont {Umarov}\ \emph {et~al.}(2008)\citenamefont {Umarov},
  \citenamefont {Tsallis},\ and\ \citenamefont {Steinberg}}]{Umarov2008}%
  \BibitemOpen
  \bibfield  {author} {\bibinfo {author} {\bibfnamefont {S.}~\bibnamefont
  {Umarov}}, \bibinfo {author} {\bibfnamefont {C.}~\bibnamefont {Tsallis}}, \
  and\ \bibinfo {author} {\bibfnamefont {S.}~\bibnamefont {Steinberg}},\ }\href
  {\doibase 10.1007/s00032-008-0087-y} {\bibfield  {journal} {\bibinfo
  {journal} {Milan journal of mathematics}\ }\textbf {\bibinfo {volume} {76}},\
  \bibinfo {pages} {307} (\bibinfo {year} {2008})}\BibitemShut {NoStop}%
\bibitem [{\citenamefont {Tirnakli}\ \emph {et~al.}(2007)\citenamefont
  {Tirnakli}, \citenamefont {Beck},\ and\ \citenamefont
  {Tsallis}}]{Tirnakli2007}%
  \BibitemOpen
  \bibfield  {author} {\bibinfo {author} {\bibfnamefont {U.}~\bibnamefont
  {Tirnakli}}, \bibinfo {author} {\bibfnamefont {C.}~\bibnamefont {Beck}}, \
  and\ \bibinfo {author} {\bibfnamefont {C.}~\bibnamefont {Tsallis}},\ }\href
  {\doibase 10.1103/PhysRevE.75.040106} {\bibfield  {journal} {\bibinfo
  {journal} {Physical Review E}\ }\textbf {\bibinfo {volume} {75}},\ \bibinfo
  {pages} {040106} (\bibinfo {year} {2007})}\BibitemShut {NoStop}%
\bibitem [{\citenamefont {Pluchino}\ \emph {et~al.}(2009)\citenamefont
  {Pluchino}, \citenamefont {Rapisarda},\ and\ \citenamefont
  {Tsallis}}]{Pluchino2009}%
  \BibitemOpen
  \bibfield  {author} {\bibinfo {author} {\bibfnamefont {A.}~\bibnamefont
  {Pluchino}}, \bibinfo {author} {\bibfnamefont {A.}~\bibnamefont {Rapisarda}},
  \ and\ \bibinfo {author} {\bibfnamefont {C.}~\bibnamefont {Tsallis}},\ }\href
  {\doibase 10.1209/0295-5075/85/60006} {\bibfield  {journal} {\bibinfo
  {journal} {{EPL} (Europhysics Letters)}\ }\textbf {\bibinfo {volume} {85}},\
  \bibinfo {pages} {60006} (\bibinfo {year} {2009})}\BibitemShut {NoStop}%
\bibitem [{\citenamefont {Pluchino}\ \emph {et~al.}(2008)\citenamefont
  {Pluchino}, \citenamefont {Rapisarda},\ and\ \citenamefont
  {Tsallis}}]{Pluchino2008}%
  \BibitemOpen
  \bibfield  {author} {\bibinfo {author} {\bibfnamefont {A.}~\bibnamefont
  {Pluchino}}, \bibinfo {author} {\bibfnamefont {A.}~\bibnamefont {Rapisarda}},
  \ and\ \bibinfo {author} {\bibfnamefont {C.}~\bibnamefont {Tsallis}},\ }\href
  {\doibase 10.1016/j.physa.2008.01.112} {\bibfield  {journal} {\bibinfo
  {journal} {Physica A: Statistical Mechanics and its Applications}\ }\textbf
  {\bibinfo {volume} {387}},\ \bibinfo {pages} {3121} (\bibinfo {year}
  {2008})}\BibitemShut {NoStop}%
\bibitem [{\citenamefont {Miritello}\ \emph
  {et~al.}(2009{\natexlab{b}})\citenamefont {Miritello}, \citenamefont
  {Pluchino},\ and\ \citenamefont {Rapisarda}}]{Miritello2009CLT}%
  \BibitemOpen
  \bibfield  {author} {\bibinfo {author} {\bibfnamefont {G.}~\bibnamefont
  {Miritello}}, \bibinfo {author} {\bibfnamefont {A.}~\bibnamefont {Pluchino}},
  \ and\ \bibinfo {author} {\bibfnamefont {A.}~\bibnamefont {Rapisarda}},\
  }\href {\doibase 10.1016/j.physa.2009.08.023} {\bibfield  {journal} {\bibinfo
   {journal} {Physica A: Statistical Mechanics and its Applications}\ }\textbf
  {\bibinfo {volume} {388}},\ \bibinfo {pages} {4818} (\bibinfo {year}
  {2009}{\natexlab{b}})}\BibitemShut {NoStop}%
\end{thebibliography}%

\end{document}